\documentclass{aastex}
\usepackage{emulateapj5}

\begin{document}

\title{THE MID-INFRARED EMITTING DUST AROUND AB AUR}
\author{C. H. Chen \& M. Jura}
\affil{Department of Physics and Astronomy, University of California,
       Los Angeles, CA 90095-1562; cchen@astro.ucla.edu; 
       jura@astro.ucla.edu}

\begin{abstract}
Using the Keck I telescope, we have obtained 11.7 $\mu$m and 18.7 $\mu$m 
images of the circumstellar dust emission from AB Aur, a Herbig Ae star. We 
find that AB Aur is probably resolved at 18.7 $\mu$m with an angular diameter 
of 1.2$\arcsec$ at a surface brightness of 3.5 Jy arcsec$^{-2}$. Most of the 
dust mass detected at millimeter wavelengths does not contribute to the 
18.7 $\mu$m emission, which is plausibly explained if the system possesses a 
relatively cold, massive disk. We find that models with an optically thick, 
geometrically thin disk, surrounded by an optically thin spherical envelope 
fit the data somewhat better than flared disk models.
\end{abstract}


\section{INTRODUCTION}
Herbig Ae/Be stars are pre-main sequence progenitors of intermediate mass A- 
and B-type stars which possess strong infrared excesses, dominated by thermal 
emission from dust grains. It is generally thought that these stars possess 
circumstellar disks (Mannings \& Sargent 1997, 2001) in which planets might 
form. The grain properties (size, temperature and composition) and the 
macroscopic geometry of the mid-infrared emitting dust around these stars are 
not yet well determined. In the simplest models, Herbig Ae/Be stars possess 
optically thick, geometrically thin disks which produce spectral energy 
distributions (SEDs) with $\nu F_{\nu}$ $\propto$ $\nu^{4/3}$, steeper than 
the observed SEDs (Hillenbrand et al. 1992). Several groups have suggested a 
variety of explanations for this discrepancy such as flared disks with a
surface layer of small grains (Chiang \& Goldreich 1997) and geometrically
thin disk embedded in spherically symmetric envelopes (Miroshnichenko et al. 
1999). We have carried out a high resolution mid-infrared imaging study of 
AB Aur to learn about the spatial distribution of warm dust which may allow 
us to distinguish among models.

AB Aur is a A0 Ve+sh (shell) Herbig Ae star with a \emph{Hipparcos} distance 
of 144 pc from the Sun. Since AB Aur lies near the zero age main sequence,
there is some uncertainty in estimating its age. Mannings \& Sargent (1997) 
estimate a total stellar luminosity and effective temperature, $L_{*}$ =
53.6 $L_{\sun}$ and $T_{eff} = 10210 K$, which corresponds to a stellar age 
and mass of $\sim$3-5 Myr and 2.5 $M_{\sun}$, using pre-main sequence tracks 
computed by D'Antona \& Mazzitelli (1994). Van den Ancker, de Winter, \& Tjin 
A Djie (1998) have estimated a total stellar luminosity and effective 
temperature, $L_{*}$ = 47 $L_{\sun}$ and $T_{eff} = 9500 K$, which corresponds
to a stellar age and mass of $\sim$2 Myr and 2.4 $M_{\sun}$, using pre-main 
sequence tracks computed by Palla \& Stahler (1993). AB Aur is an excellent 
source for high resolution mid-infrared imaging because it is bright and close
by. It is one of four stars in the Hillenbrand et al. (1992) sample which 
possess a \emph{Q} band magnitude brighter than -1.0 and lies at a distance 
less than 150 pc. All models for the dust distribution around AB Aur presume 
that it is a single star. If AB Aur is a binary, then current models are 
incomplete.

Millimeter aperture synthesis imaging of AB Aur has resolved a rotating disk 
of $^{13}$CO (J=1$\rightarrow$0), with a radius of $\sim$450 AU and at a 
position angle of 79$\arcdeg$, around AB Aur, with velocities consistent with
Keplerian rotation (Mannings \& Sargent 1997). Based upon the aspect ratio of 
the major:minor axes, Mannings \& Sargent (1997) find an inclination of 
76\arcdeg \ for the disk. Previous mid-infrared imaging studies of AB Aur 
report spatially resolved observations of the mid-infrared emission at 
17.9 $\mu$m using the 5 m Hale telescope (Marsh et al. 1995). They find an 
elongated disk, with an East-West extent of 80$\pm$20 AU and a North-South 
extent of 25$\pm$20 AU and thus a position angle consistent with that 
observed for the CO disk. However, subsequent higher resolution near infrared 
interferometry (Millan-Gabet et al. 1999; Millan-Gabet, Schloerb, \& Traub 
2001) and scattered light imaging of AB Aur (Grady et al. 1999) suggest a 
disk inclination of less than 45$\arcdeg$. 

The dust around AB Aur has been studied with a variety of techniques. Near 
infrared, interferometric observations at \emph{H} and \emph{K'} bands have 
resolved a ring of emission with a radius of 0.35 AU (Millan-Gabet et al. 
1999). \emph{ISO} 2-200 $\mu$m spectra have revealed the presence of PAH, FeO,
and olivine emission features and Si-O and O-Si-O bending modes (van den 
Ancker et al. 2000). However, to date, no complete view of the dust around AB 
Aur has been constructed. Thus, we discuss models for the dust around AB Aur 
in the context of millimeter continuum fluxes, infrared spectra, and near 
infrared visibilities. In addition, we compare models for the dust around AB 
Aur with our high resolution Keck 18.7 $\mu$m map.  

\section{OBSERVATIONS}
Our data were obtained on 2000 February 20 (UT) and 2000 August 9 (UT) at the 
Keck I telescope using the Long Wavelength Spectrometer (LWS) which was built 
by a team led by B. Jones and is described on the Keck web page. The LWS is a 
128$\times$128 SiAs BIB array with a pixel scale at the Keck telescope of 
0.08$\arcsec$ and a total field of view of 10.2$\arcsec\times$10.2$\arcsec$. We
used the ``chop-nod'' mode of observing and two different filters: 11.2-12.2 
$\micron$ and 18.2-19.2 $\mu$m. The seeing on 2000 August 9 was significantly 
better than that on 2000 February 20; thus, we concentrate our analysis on the 
data obtained in August 2000. We used $\alpha$ Cet for flux and point spread
function (PSF) calibrations and MWC 480 (an unresolved Herbig Ae star) and 
$\gamma$ And for additional PSF calibrations. The data were reduced at UCLA 
using standard LWS routines.

We flux calibrate our February and August data using the result for 
$\alpha$ Tau and $\alpha$ Cet that $F_{\nu}(11.5 \mu m)$ = 507.8 Jy and 
174.5 Jy, respectively and $F_{\nu}(19.3 \mu m)$ = 189.4 Jy and 66.3 Jy 
(Gezari et al. 1987) extrapolated to our bands assuming that $F_{\nu} 
\propto \nu^{2}$ between 10 and 20 $\mu$m. For AB Aur, we find 
$F_{\nu}(11.7 \mu m)$ = 25$\pm$2 Jy and 19$\pm$2 Jy for February and August,
respectively, and $F_{\nu}(18.7 \mu m)$ = 31$\pm$2 Jy and 17$\pm$2 Jy for
February and August, respectively. We conservatively estimate the 
uncertainties associated with our measurements from the drift in the fluxes of
all of our standard stars throughout the night. Comparison of our fluxes and 
of published fluxes in the 10 $\mu$m and 20 $\mu$m bands (listed in Tables 2 
and 3) suggest that AB Aur is variable at mid-infrared wavelengths. The
24$\pm$10\% 11.7 $\mu$m variability we observe is consistent with the 
21.7$\pm$6.6\% IRAS 12 $\mu$m 6 month variability reported by Prusti \& 
Mitskevich (1994). The 46$\pm$8\% 18.7 $\mu$m variability we observe is 
significantly larger than the 14.3$\pm$8.5\% variability IRAS 25 $\mu$m 
6 month variability reported by Prusti \& Mitskevich (1994). However, the 
absolute magnitudes of our 18.7 $\mu$m fluxes are similar to 17 $\mu$m fluxes 
measured using \emph{ISO} and the IRTF, 24.4 Jy (Thi et al. 2001) and 
22.7 Jy (Richter et al. 2002) respectively. 

The color temperature, inferred from our 11.7 $\mu$m and 18.7 $\mu$m 
photometry, for our data agrees within the error bars for our two epochs
of observations. The flux ratio, 
$F_{\nu}$(11.7 $\mu$m)/$F_{\nu}$(18.7 $\mu$m), is 0.8$\pm$0.1 and 1.1$\pm$0.2 
for Februrary 2000 and August 2000, respectively. Thus, we can not rule out
the possibility that changes in intrinsic luminosity of the star generate the
observed variability. AB Aur possesses strong variability in the H$\alpha$ 
P Cygni profile and in measurements of the linear polarization of UBVRI. 
Based upon monitoring of these diagnostics in 1993 and 1994, Beskrovnaya et 
al. (1995) suggest that circumstellar inhomogeneities may be responsible for
the variability on hour to month long timescales. Prusti \& Mitskevich (1994)
suggest that dust formation in the stellar wind or dust on eccentric orbits 
may be reponsible for the mid-infrared variability.

We generate images from our chop-nod sets by coadding chop pairs from both
nod positions in each of our nod sets. The 18.7 $\mu$m PSF contains a 
ghost and diffuse asymmetric scattering while the 11.7 $\mu$m PSF is clean.  
Although the seeing was excellent, the PSFs varied somewhat during the night.
We show line cuts through AB Aur and our 11.7 $\mu$m standards in Figure 1a 
and through AB Aur and our 18.7 $\mu$m standards perpendicular to and parallel
to the scattered light in Figures 1b and 1c. The Airy rings can be seen in all
of the 18.7 $\mu$m line cuts and half of the 11.7 $\mu$m line cuts. Since 
there is some variation in the PSF at 18.7 $\mu$m, we compare our images of 
AB Aur to 3 PSF calibrators: $\alpha$ Cet, observed 57 minutes earlier, 
MWC 480 (an unresolved Herbig Ae star), observed 11 minutes later, and 
$\gamma$ And, observed 2 hours, 26 minutes earlier than AB Aur. We use MWC 480
as our primary PSF because it was observed closest in time to AB Aur. We also 
use $\gamma$ And because it was observed with the worst seeing during the 
evening.

AB Aur is probably resolved at 18.7 $\mu$m with an angular diameter of 
1.2$\arcsec$ at a surface brightness of 3.5 Jy arcsec$^{-2}$ (10\% of the
maximum surface brightness). An image of our calibrator $\alpha$ Cet at 
18.7 $\mu$m is shown in Figure 2 with contours showing 1\%, 2\%, 3\%, 5\%, 
10\%, 25\%, 50\%, and 99\% of the maximum surface brightness. We display
$\alpha$ Cet instead of MWC 480 because it is a significantly bright object,
making the ghost and diffuse scattered light significantly easier to see. An 
image of AB Aur at 18.7 $\mu$m is shown in Figure 3 for comparison with the 
same contours as in the image of $\alpha$ Cet. The diameter of AB Aur at 10\% 
of the maximum surface brightness is 50\% larger than the same contour in the 
image of $\alpha$ Cet. In addition, the lowest surface brightness contours 
(1\%, 2\%, 3\%, and 5\%) are more symmetric in the image of AB Aur than in 
the image of $\alpha$ Cet. Our image of AB Aur at 18.7 $\mu$m does not 
exhibit the same asymmetry reported by Marsh et al. (1995) at 17.9 $\mu$m. 
AB Aur appears unresolved at 11.7 $\mu$m, with a Full Width at Half Maximum 
(FWHM) of 0.30$\arcsec$, consistent with a point source.  

\section{THE CIRCUMSTELLAR DISK AND MORE?}
The total mass of dust around AB Aur has been estimated from millimeter
contiuum measurements in an elliptical beam with FWHM 3.5$\arcsec$ $\times$ 
5.6$\arcsec$. At $\lambda$ = 2.7 mm, Mannings \& Sargent (1997) find that AB 
Aur appears unresolved and measured a flux, $F_{\nu}$(2.7 mm) = 10.6 $\pm$ 0.4 
mJy which, in their models, corresponds to $M_{dust}$ = 2.0 $\times$ 10$^{29}$ 
g.

We can estimate the minimum mass of warm dust contributing to the 18.7
$\mu$m flux. If the dust is optically thin and the population has a single
temperature, then 
\begin{equation}
M_{warm} = \frac{F_{\nu} D_{*}^2}{B_{\nu}(T_{gr}) \chi_{\nu}}
\end{equation}
(Jura et al. 1995) where $\chi_{\nu}$ is the dust absorption opacity and 
$D_{*}$ is the distance to the object. We estimate the temperature of the 
dust, $T_{gr}$, from the ratio of our 11.7 $\mu$m and 18.7 $\mu$m fluxes, 
assuming a single population of grains. If $T_{gr}$ = 250 K, 
$F_{\nu}$(18.7 $\mu$m) = 17.2 Jy, and $\chi_{\nu}$ = 1000 cm$^{2}$/g 
(Ossenkopf, Henning, \& Mathis 1992), then $M_{warm}$ = 1.2 $\times$ 
10$^{25}$ g, substantially less than that inferred from millimeter 
continuum measurements. Therefore, a large amount of cold dust must exist in 
the system, presumably in a disk. 

\section{COMPARISON WITH MODELS}
We use our data to evaluate the following four detailed models for the dust 
around AB Aur: (1) An optically thick, geometrically thin disk surrounded by 
an optically thin, spherically symmetric envelope (Vinkovic et al. 2002). 
(2) Two spherically symmetric dust shells, whose grain properties 
(composition, size and spatial distribution) are inferred from \emph{ISO} 
2 - 200 $\mu$m spectra (Bouwman et al. 2000). (3) A geometrically thin, flared
disk in vertical hydrostatic equilibrium (D'Alessio et al. 1998). (4) A flared
disk with an optically thin surface layer of small grains and an infrared 
emitting rim (Dullemond, Domink, \& Natta 2001).

We convolve these models for the predicted surface brightness of AB Aur at 
18.7 $\mu$m with the PSF and compare them with our observations of AB Aur. 
Each model was interpolated to provide surface brightnesses on the same grid 
as the observations. Since line cuts through each of the models, convolved 
with a PSF, compared with the AB Aur data look similar independent of position
angle, we only show line cuts along the North-South direction of each of the 
models, convolved with a PSF, compared with the AB Aur data. In Figure 4a, 
MWC 480, an unresolved Herbig Ae star observed closest in time to AB Aur, was 
used as the PSF. In Figure 4b, $\gamma$ And, the most extended PSF observed 
during the night, was used. Line cuts through different position angles appear
similar; thus, we show line cuts through only one orientation. In addition, we
calculate the reduced $\chi^{2}$ for the inner 1.2$\arcsec$ $\times$ 
1.2$\arcsec$ centered on AB Aur, using MWC 480 as the PSF. In both analyses, 
we scale the image of the model, convolved with the PSF, so that the maximum 
surface brightness is equal to the measured maximum surface brightness of 
AB Aur. Since AB Aur is located at a distance of 144 pc, the flux expected 
from the stellar photosphere at 18.7 $\mu$m, $F_{\nu}(18.7 \mu m)$ = 0.032 Jy,
is significantly less than the smallest measured excess flux, 
$F_{\nu}(18.7 \mu m)$ = 17.2 Jy, and can be neglected. We calculate the 
reduced $\chi^{2}$ for each model assuming that the uncertainty in the surface
brightness is ~10\%, approximately the uncertainty in the 18.7 $\mu$m 
photometry. Finally, we also compare the models to measured near infrared 
interferometric visibilities, infrared spectra, and millimeter continuum 
measurements.

\subsection{A Spherically Symmetric Envelope?}
Vinkovic et al. (2002) have modeled the SED of AB Aur with a system which
consists of a star with radius, $R_{*}$, and effective temperature, $T_{*}$, 
surrounded by a geometrically thin and optically thick passive disk extending 
from $R_{*}$ to an outer radius $R_{disk}$ = 240 AU. The optically thick disk 
is assumed to have an inclination, $i$ = 76$\arcdeg$, consistent with the CO 
observations of Mannings \& Sargent (1997). They also incorporate a  spherical
dust envelope with an inner radius, $R_{sub}$ = 99.3 $R_{*}$ and an outer 
radius, 1000 $R_{sub}$. At distances less than $R_{sub}$, grains in the 
envelope are sublimated. The mass density and grain size distributions are 
given by
\begin{equation}
\rho(r) = \rho_{o} \left[  \left( \frac{R_{sub}}{r} \right)^{2} 
                         + \frac{0.2 R_{sub}}{r} \right]
\end{equation}
\begin{equation}
n(a) = A a^{-3.5}
\end{equation}
where $r$ is the distance between the star and the circumstellar material and
$a$ is the grain radius. The dust grains are assumed to be an interstellar 
mixture of 53\% silicates and 47\% graphite by number (Draine \& Lee 1984) 
with a minimum grain radius, $a_{min}$ = 0.005 $\mu$m, and a maximum grain 
radius $a_{max}$ = 0.250 $\mu$m. They assume an optically thick, geometrically
thin (flat) disk with an outer temperature of 25 K, a disk mass, $M_{disk}$ = 
8.6 $\times$ 10$^{30}$ g, and an envelope mass, 
$M_{env}$ = 2.8 $\times$ 10$^{27}$ g, corresponding to an optical depth, 
$\tau_{V}$ = 0.44, consistent with the value measured by van den Ancker et al.
(2000) ($A_{V}$ = 0.5). Their best fit model for the SED for AB Aur yields a 
total flux $F_{\nu}(18.7 \mu m)$ = 25 Jy with 12 Jy from the disk and 13 Jy 
from the envelope. 

We compare the inner 1.2$\arcsec$ $\times$ 1.2$\arcsec$ of our 18.7 $\mu$m map
of AB Aur with the model (Vinkovic et al. 2002). When the intensity map for 
this model is convolved with the PSF and scaled so that the maximum surface 
brightness is equal to that measured for AB Aur, the reduced $\chi^2$ = 4.6.
Their model agrees well with our data.  

However, the Vinkovic et al. (2002) model does have some limitations. They 
do not provide a physical explanation for the stability of the envelope. In
addition, their model does not attempt to reproduce the detailed infrared 
spectra; a standard interstellar grain mixture is assumed. While the near 
infrared fluxes are fit, the measured visibilities are not considered. 
Finally, the model assumes a dust disk mass which is more than an order of 
magnitude higher than is inferred from millimeter continuum measurements.

\subsection{Modeling the Infrared Spectra}
Bouwman et al. (2000) infer dust composition, spatial distribution, and  grain
sizes by fitting \emph{ISO} 2 - 200 $\mu$m spectra. For simplicity, they 
assume that the dust around AB Aur is optically thin and contained within 
spherically symmetric shells and neglect the presence of a circumstellar disk.
They describe the density distribution, $\rho(r)$, and the grain size 
distribution, $n(a)$, with power laws.
\begin{equation}
\rho(r) = \rho_{o} \left( \frac{R_{in}}{r} \right)
\end{equation}
\begin{equation}
n(a) = A \left( \frac{a_{min}}{a} \right)^{m}
\end{equation}
where $\rho_{o}$ is the dust density at the inner edge of the dust shell,
$R_{in}$, and $a_{min}$ is the minimum dust grain size. They find that a 
combination of two populations of dust, a ``hot'' population at 1 - 11 AU, 
containing 5.2 $\times$ 10$^{24}$ g (m = 2.8) and a ``cold'' 
population at 28 - 175 AU, containing 6.6 $\times$ 10$^{28}$ g 
(m = 2.0), reproduces the observed 2-200 $\mu$m \emph{ISO} spectrum well. 
Their model fit parameters, including grain composition and the mass fraction 
of dust in each species, are reproduced in Table 4. Bouwman et al. (2000)
assume that the iron oxide grains are small (2$\pi a$/$\lambda$ $<<$ 1) and
do not infer a size distribution for this species.

Since Bouwman et al. (2000) do not predict the surface brightness of their
model, we estimate the surface brightness of optically thin, spherically 
symmetric dust shells as a function of angular offset on the sky, assuming 
the same ``hot'' and ``cold'' populations of dust with the same power law 
dependences for $\rho(r)$ and $n(a)$ and the same composition. The specific 
intensity at an offset, $b$, from the star is calculated by integrating along 
the line of sight through the envelope, perpendicular to the sky. 
\begin{equation}
I_{\nu} = \sum_{all \ species} \left[ \int_{x_{min}}^{x_{max}} 
          \rho_{i}(r) \int_{a_{min,i}}^{a_{max,i}} B_{\nu}[T_{gr,i}(x)] 
          n_{i}(a) Q_{abs,i} \pi a^{2} da dx \right]
\end{equation}
where $r = \sqrt{b^2 + x^2}$ is the distance between the star and the dust 
grains whose radiation is summed in the integral. For the cold grains,
$x_{max}$ = $\sqrt{\textup{175 AU}^{2} - b^{2}}$ and $x_{min}$ = 0 for $b$ 
$<$ 28 AU and $x_{min}$ = $\sqrt{\textup{28 AU}^{2} - b^{2}}$ for $b$ $\geq$ 
28 AU. For the warm grains,$x_{max}$ = $\sqrt{\textup{11 AU}^{2} - b^{2}}$ and
$x_{min}$ = 0 for $b$ $<$ 1 AU and $x_{min}$ = $\sqrt{\textup{1 AU}^{2} - 
b^{2}}$ for $b$ $\geq$ 1 AU. The grain temperature for each species, $T_{gr}$, 
is calculated numerically, at a distance $r$ from the star, assuming that the 
grains are in radiative equilibrium.
\begin{equation}
\left( \frac{R_{*}}{D} \right)^{2} \int_{0}^{\infty} 
    Q_{abs}(\nu) B_{\nu}(T_{*}) d\nu
= 4 \int_{0}^{\infty} Q_{abs}(\nu) B_{\nu}(T_{gr}) d\nu
\end{equation}
where $T_{*}$ and $R_{*}$ are the stellar radius and temperature, $D$ is
the distance between the star and the dust grain. We calculate the absorption 
coeffecient for each species of dust, $Q_{abs}(\nu)$, and wavelengths less 
than 0.2 $\mu$m following Bouwman et al. (2000). We calculate a total flux 
$F_{\nu}(18.7 \mu m)$ = 45 Jy, with 20 Jy produced by the hot population and 
25 Jy produced by the cold population.

The Bouwman et al. (2000) model does not fit our data particularly well.
After scaling the surface brightness to that of AB Aur, convolving a 
1.2$\arcsec$ $\times$ 1.2$\arcsec$ model of the dust emission with the PSF 
(MWC 480) yields a reduced $\chi^2$ = 8.6.

Bouwman et al. (2000) use grain materials which reproduce the \emph{ISO} 2 - 
200 $\mu$m emission features well. They have some difficulty reproducing the 
observed near infrared visibilities. Their best fit model includes iron grains
with radii 0.01 - 0.1 $\mu$m to produce the near infrared emission close to 
the star. Iron is chosen as a material because it is one of the few materials 
which is likely to survive at temperatures $\sim$1500 K and its emissivity is 
high at near infrared wavelengths and low at mid- and far infrared 
wavelengths; there is no direct evidence for the presence of metallic iron 
from emission line features. The Bouwman et al. (2000) model is inconsistent 
with the measured visibility curves. A single grain population of metallic 
iron grains with radius, $a$ = 0.32 $\mu$m, and minimum distance, $R_{min}$ = 
0.35 AU fits the visibilities the best and reproduces the measured near 
infrared size of AB Aur. Bouwman et al. (2000) neglect the presence of a 
circumstellar disk altogether.

\subsection{A Flared Disk in Hydrostatic Equilibrium?}
Another possibility is that the flat SED is produced by a flared disk. Flared 
disks intercept more stellar flux than geometrically flat disks, allowing 
surface grains to be warmed to higher temperatures. Flared disks have been 
observed around the young stellar objects HH 30 and HK Tau/c in scattered 
light using \emph{HST} (Burrows et al. 1996; Stapelfeldt 1998). Simple 
semi-analytical models of flared disks with a layer of small 
(2$\pi a$/$\lambda$ $<<$ 1), optically thin, surface grains can reproduce the 
SEDs of some T Tauri and Herbig Ae stars (Chiang \& Goldreich 1997; Chiang et 
al. 2001). Detailed numerical modeling has shown that stellar radiation is 
capable of heating the outer regions of the disk so that they become optically
thin and vertically isothermal (D'Alessio et al. 1998). D'Alessio, Calvet, \& 
Hartmann (2001) successfully model the SEDs and images of edge-on disks like
HH30 and HK Tau/c. 

D'Alessio et al. (2002) model the intensity of the dust around AB Aur at 
18.7 $\mu$m assuming that the central star has mass, $M_{*}$ = 2.4 $M_{\sun}$,
radius, $R_{*}$ = 2.5 $R_{\sun}$, and effective temperature $T_{*}$ = 9500 K 
and is surrounded by a disk with inner radius, $R_{min}$ = 3 $R_{*}$ and an 
outer radius, $R_{disk}$ = 200 AU. They assume a disk inclination, 
$i$ = 75.5$\arcdeg$, and a disk dust mass, $M_{dust}$ = 5.6 $\times$ 10$^{29}$
g, with a gas:dust ratio of 100:1. They assume that the disk is composed of 
silicates, trolite, ice, and organics with mass fractional abundances 
(relative to the gas mass) of 0.0034, 0.000768, 0.0056, and 0.0041 
respectively (D'Alessio et al. 2001). The grains are assumed to have a size 
distribution described by a power law, 
\begin{equation}
n(a) = A a^{-3.5}
\end{equation}
with a minimum grain radius, $a_{min}$ = 0.005 $\mu$m and a maximum grain 
radius, $a_{max}$ = 1 mm. The assumption of grain growth in this model changes
the opacity of the dust grains at small and large wavelengths and therefore 
the temperature distribution of the disk. Larger grain sizes decrease the 
fraction of stellar radiation intercepted by the disk resulting in a colder 
disk. The model assumes an accretion rate, \.{M} = 1.0 $\times$ 10$^{-8}$ 
$M_{\sun}$/year, corresponding to a accretion luminosity of 2 $L_{\sun}$. 

The flared disk model does not fit our data particularly well. It predicts too
little 18.7 $\mu$m flux, 5 Jy, compared with all of the 20 $\mu$m band flux
measurements. This may partly be the result of the high assumed inclination,
$i$ = 75.5$\arcdeg$. Recent near infrared interferometry (Millan-Gabet et al. 
1999) and scattered light imaging (Grady et al. 1999) suggests that the 
inclination of the AB Aur disk is probably less than 45$\arcdeg$. Since the 
luminosity of the disk depends directly on the cosine of the inclination, the 
estimated flux of the D'Alessio disk might be as much as 2.8 times higher if 
it is calculated for an inclination of 45$\arcdeg$. When the maximum surface 
brightness is scaled to that of AB Aur, convolving a model of the inner 
1.2$\arcsec$ $\times$ 1.2$\arcsec$, with the most likely PSF (MWC 480) yields 
a reduced $\chi^2$ = 9.2. However, convolving the model with the worst PSF 
($\gamma$ And), yields a model which appears consistent with the observations.

The D'Alessio et al. (2002) model has some limitations. They assume a Pollack 
et al. (1994) grain mixture and ignore detailed grain composition information 
from infrared spectra. They do not fit the observed near infrared 
visibilities. However, their assumed circumstellar dust disk mass is an 
approximate agreement with that inferred from millimeter continuum 
measurements. 

\subsection{A Flared Disk with a Hot Inner Rim?}
The simpled flared disk model with a surface layer of small, optically thin 
grains predicts too little disk emission for AB Aur at near infrared 
wavelengths. Dullemond et al. (2001) have refined the Chiang \& Goldreich 
(1997) flared disk model to allow for direct stellar heating of dust grains at
the inner disk radius. In their model, grains sublime when their temperatures 
reach $T_{rim}$ (= $T_{sub}$ = 1500 K). The direct stellar radiation on the 
inner rim surface warms the grains effectively, creating the observed near 
infrared excess, puffing up the inner disk, and casting a shadow on adjacent 
portions of the disk. Dullemond et al. (2001) model the AB Aur system with a 
star, with $M_{*}$ = 2.4 $M_{\sun}$, $R_{*}$ = 2.5 $R_{\sun}$, $T_{eff}$ = 
9520 K, ($L_{*}$ = 47 $L_{\sun}$) surrounded by a disk with an inner rim
$R_{rim}$ = 0.52 AU away from the star and an outer radius $R_{disk}$ = 
400 AU. The vertical height of the inner rim is $H_{rim}$ = 0.19 $R_{rim}$. 
In modeling the SED, the circumstellar disk is assumed to have an inclination,
$i$ = 65$\arcdeg$, less than that inferred from the CO observations of 
Mannings \& Sargent (1997), more consistent with the 45$\arcdeg$ inclination 
inferred from scattered light observations of Grady et al. (1999). However,
the model of intensity as a function of position was calculated assuming that
the AB Aur disk was face-on. Dullemond et al. (2001) assume a disk surface 
density 
\begin{equation}
\Sigma = \Sigma_{o} \left( \frac{r}{1 \ \textup{AU}} \right)^{-2}
\end{equation}
where $\Sigma_{o}$ = 10$^{4}$ g/cm$^{2}$ with a total mass of 4.0 $\times$ 
10$^{29}$ g. The gas:dust ratio is assumed be 100:1. The dust grains are 
assumed to be a mixture of 95\% astronomical silcate (Draine \& Lee 1984) and 
5\% amorphous carbon (Ivezic et al. 1997). The rim and disk produce an 18.7 
$\mu$m flux, $F_{\nu}(18.7 \mu m)$ = 49 Jy. 

The Dullemond et al. (2001) model does not agree with our mid-infrared 
observations. The reduced $\chi^{2}$ for the model convolved with the PSF 
(MWC 480), in the inner 1.2$\arcsec$ $\times$ 1.2$\arcsec$, is 84, when 
the convolved image is scaled to have the same maximum surface brightness 
as AB Aur. Since the rim produces an order of magnitude less emission than 
the disk at 18.7 $\mu$m, the Dullemond et al. (2001) model is essentially a 
Chiang \& Goldreich (1997) flared disk with an inner radius of 8 AU. If the 
disk were to extend to an inner radius of 0.5 AU, 8\% more 18.7 $\mu$m 
emission would be produced (Dullemond et al. 2001) in the central pixel, 
only slightly affecting our normalization. Thus, the Chiang \& Goldreich 
(1997) model similarly does not agree with our mid-infrared observations.

Dullemond et al. (2001) do fit other observational data well. They reproduce 
the measured near infrared fluxes and visibilities by placing the hot inner 
rim at the grain sublimation distance. Their assumed disk mass agrees with 
the dust mass, inferred from millimeter continuum measurements. They use a
silicate/carbon grain mixture and do not match the infrared spectrum.

\section{DISCUSSION}
Models for the dust around AB Aur must fit a suite of multi wavelength 
observations if they are to be successful. In Table 5, we summarize the 
current ability of models to fit various observational data sets. In Table 6,
we compare the predicted 18.7 $\mu$m fluxes, dust disk masses, and reduced 
$\chi^{2}$s of each model to those inferred from observations. By examining 
the line cuts shown in Figures 4a and 4b, we see that the Dullemond et al. 
(2001) model for the dust intensity predicts too much flux at distances 
$\geq$0.24$\arcsec$ (35 AU) from the star. If MWC 480 more accurately 
represents the PSF, then the Bouwman et al. (2000) model predicts too much 
emission and the D'Alessio (2002) model predicts too little emission at 
distances $\geq$0.40$\arcsec$ (58 AU) from the star. In this case, the 
Vinkovic et al. (2002) model is favoured. Flared disk models are not able to 
reproduce both the compact size of AB Aur and the strong mid-infrared flux. 
However, there is some uncertainty in the seeing at the time that AB Aur was 
observed; thus, it is not possible to rule out the flared disk model. 
Limitations exist with all the models, as described above. Future models of 
AB Aur must consider the grain composition, if they are to accurately model 
the SED and dust emission.

\section{CONCLUSIONS}
We have obtained high resolution mid-infrared images of AB Aur at 11.7 $\mu$m
and 18.7 $\mu$m using the LWS on the Keck I telescope.

1. AB Aur is probably resolved at 18.7 $\mu$m, with an angular diameter of 
1.2$\arcsec$ at a surface brightness of 3.5 Jy arcsec$^{-2}$, and unresolved
at 11.7 $\mu$m, with a FWHM = 0.30$\arcsec$.

2. Only $\sim$10$^{25}$ g of dust is required to produce the 18.7 $\mu$m
emission. Millimeter interferometry indicates the presence of 2 $\times$ 
10$^{29}$ g of dust. Thus, a large amount of cold dust must exist in the 
system, presumably in a disk. 

3. The spatial distribution of the 18.7 $\mu$m emission is most consistent 
with radiation from an optically thick, geometrically thin disk embedded in a 
spherically symmetric dust envelope (Vinkovic et al. 2002); however, the 
presence of a flared disk can not be excluded. None of the models yet 
reproduces all of the available observations.

\acknowledgements

This work has been supported by funding from NASA. We thank A. Sargent, 
B. Zuckerman, and our referee J. Bouwman for their comments. We also thank
P. D'Alessio, C. Dullemond, and D. Vinkovic for providing us with models 
for the intensity of dust emission around AB Aur. P. Plavchan provided an 
IDL code to display the mid-infrared excess from AB Aur.

\begin{deluxetable}{lll}
\singlespace
\tablecaption{AB Aur Properties} 
\tablehead{
    \colhead{Quantity} &
    \colhead{Adopted Value} &
    \colhead{Reference} \\
}
\tablewidth{0pt}
\tablecolumns{3}
\startdata
    Spectral Type & A0Ve+sh & 1 \\
    Distance & 144 pc & 2 \\
    Effective Temperature (T$_{eff}$) & 9500-10210 K & 2, 5 \\
    Stellar Radius (R$_{*}$) & 2.3-2.5 R$_{\sun}$ & 2, 5\\
    Stellar Luminosity (L$_{*}$) & 47-53.6 L$_{\sun}$ & 2, 5 \\
    Stellar Mass (M$_{*}$) & 2.4-2.5 M$_{\sun}$ & 2, 5 \\
    Rotational Velocity ($v\sin i$) & 80 km/sec & 3\\
    Fractional Dust Luminosity & 0.44 & 4 \\
    \ \ \ \ \ ($L_{IR}/L_{*}$) & & \\ 
    Estimated Age & 2-5 Myr & 2, 5 \\
\enddata
\tablerefs{(1) B\"{o}hm \& Catala (1993);
           (2) van den Ancker et al. (2000);
           (3) B\"{o}hm \& Catala (1995);
           (4) Natta et al. (2001);
           (5) Mannings \& Sargent (1997)
          }
\end{deluxetable}

\begin{deluxetable}{cccc}
\singlespace
\tablecaption{Measured 10 $\mu$m Fluxes}
\tablehead{
    \colhead{Date} & 
    \colhead{Wavelength} & 
    \colhead{Flux} & 
    \colhead{Reference} \\
    \colhead{Observed} & 
    \colhead{($\mu$m)} & 
    \colhead{(Jy)} & 
    \omit \\
}
\tablewidth{0pt}
\tablecolumns{4}
\startdata
     & 10.1 & 25.2 & Hillenbrand et al. (1992)\\ 
    1993 Oct 23 & 11.7 & 20.6$\pm$0.6 & Marsh et al. (1995)\\ 
    1994 Nov 14 & 11.7 & 23.0$\pm$0.7 & Marsh et al. (1995)\\ 
    2000 Feb 20 & 11.7 & 25$\pm$2 & this paper\\ 
    2000 Aug 9 & 11.7 & 19$\pm$2 & this paper\\      
     & 12 & 30.4$\pm$1.8 & Prusti \& Mitskevich (1994) \\
     & 12 & 23.8$\pm$1.4 & Prusti \& Mitskevich (1994) \\
     & 12 & 18.7$\pm$1.3 & Richter et al. (2002) \\
\enddata
\end{deluxetable}

\begin{deluxetable}{cccc}
\singlespace
\tablecaption{Measured 20 $\mu$m Fluxes}
\tablehead{
    \colhead{Date} & 
    \colhead{Wavelength} & 
    \colhead{Flux} & 
    \colhead{Reference} \\
    \colhead{Observed} & 
    \colhead{($\mu$m)} & 
    \colhead{(Jy)} & 
    \omit \\ 
}
\tablewidth{0pt}
\tablecolumns{4}
\startdata 
     & 17 & 24.4 & Thi et al. (2001) \\
     & 17 & 22.7$\pm$2.3 & Richter et al. (2002) \\
    1994 Nov 14 & 17.9 & 34.7$\pm$5.2 & Marsh et al. (1995) \\ 
    2000 Feb 20 & 18.7 & 31$\pm$3 & this paper\\ 
    2000 Aug 9 & 18.7 & 17$\pm$2 & this paper\\ 
     & 20.25 & 27.2 & Hillenbrand et al. (1992) \\ 
     & 25 & 49.6 & Prusti \& Mitskevich (1994) \\
     & 25 & 42.5 & Prusti \& Mitskevich (1994) \\ \hline
\enddata
\end{deluxetable}

\begin{deluxetable}{lcccc}
\singlespace
\tablecaption{Grain Properties}
\tablewidth{0pt}
\tablecolumns{5}
\startdata
  Vinkovic et al. (2002): & & & & \\
   & $M_{frac}$ & $a$ ($\mu$m) & & \\
  Astronomical Silicate & 0.63 & 0.005-0.250 \\
  Carbon & 0.37 & 0.005-0.250 & & \\ \hline
  Bouwman et al. (2000): & & & & \\  
   & Hot $M_{frac}$ & Hot $a$ ($\mu$m) & Cold $M_{frac}$ & 
      Cold $a$ ($\mu$m)\\ 
    Olivine & 0.71 & 0.01-5.0 & 0.74 & 0.01-126 \\
  Carbon & 0.16 & 0.01-2.0 & 0.11 & 0.01-32 \\
  Water Ice & - & - & 0.15 & 0.1-40.0 \\
  Iron & 0.06 & 0.01-1.0 & - & - \\
  Iron Oxide & 0.07 & - & 9$\times$10$^{-4}$ & - \\ \hline
  D'Alessio (2002): & & & & \\
    & $M_{frac}$ & $a$ ($\mu$m) & & \\
  Silicate & 0.25 & 0.005-1000 & & \\
  Trolite & 0.05 & 0.005-1000 & & \\
  Ice & 0.40 & 0.005-1000 & & \\
  Organics & 0.30 & 0.005-1000 & & \\ \hline
  Dullemond et al. (2001): & & & & \\
    & $M_{frac}$ & $a$ ($\mu$m) & & \\
  Astronomical Silicate & 0.95 & 0.005-0.250 & & \\
  Amorphous Carbon & 0.05 & 0.16 & & \\ \hline
\enddata
\end{deluxetable}

\begin{deluxetable}{lccc}
\singlespace
\tablecaption{Comparing the Models with Observations} 
\tablehead{
    \omit &
    \colhead{Infrared} &
    \colhead{18.7 $\mu$m} &
    \colhead{Near infrared} \\
    \omit &
    \colhead{Spectrum} &
    \colhead{Image} &
    \colhead{Visibilities} \\
}
\tablewidth{0pt}
\tablecolumns{4}
\startdata
    Vinkovic et al. (2002) & Omitted & Yes & Omitted \\
    Bouwman et al. (2000) & Yes & No & No \\
    D'Alessio (2002) & Omitted & Possibly & Omitted \\
    Dullemond et al. (2001) & Omitted & No & Yes \\
\enddata
\end{deluxetable}

\begin{deluxetable}{lccc}
\singlespace
\tablecaption{Model Properties}
\tablehead{
    \omit &
    \colhead{$F_{\nu}$(18.7 $\mu$m)} & 
    \colhead{Dust Mass} & 
    \colhead{Reduced} \\
    \omit & 
    \colhead{(Jy)} & 
    \colhead{(g)} & 
    \colhead{$\chi^{2}$} \\
}
\tablewidth{0pt}
\tablecolumns{4}
\startdata
    Observed Quantities & 31, 17 & 2 $\times$ 10$^{29}$ 
     \tablenotemark{\dagger} & \\ \hline
    Predicted Quantities: & & & \\
    \ \ \ Vinkovic et al. (2002) & 25 & 8.6 $\times$ 10$^{30}$ &  4.6 \\
    \ \ \ Bouwman et al. (2000) & 45 & 6.6 $\times$ 10$^{28}$
     \tablenotemark{\ddagger} & 8.6 \\
    \ \ \ D'Alessio (2002) & 5 & 5.6 $\times$ 10$^{29}$ & 9.2 \\
    \ \ \ Dullemond et al. (2001) & 49 & 4.2 $\times$ 10$^{29}$ & 84\\ \hline
\enddata
\tablenotetext{\dagger}{Disk mass estimated from 2.7 mm interferometry 
    (Mannings \& Sargent 1997)}
\tablenotetext{\ddagger}{Dust in the Bouwman et al. (2000) model is located 
    in spherically symmetric dust shells and not in a circumstellar disk.}
\end{deluxetable}

\begin{figure}[ht]
\figurenum{1}
\epsscale{0.45}
\plotone{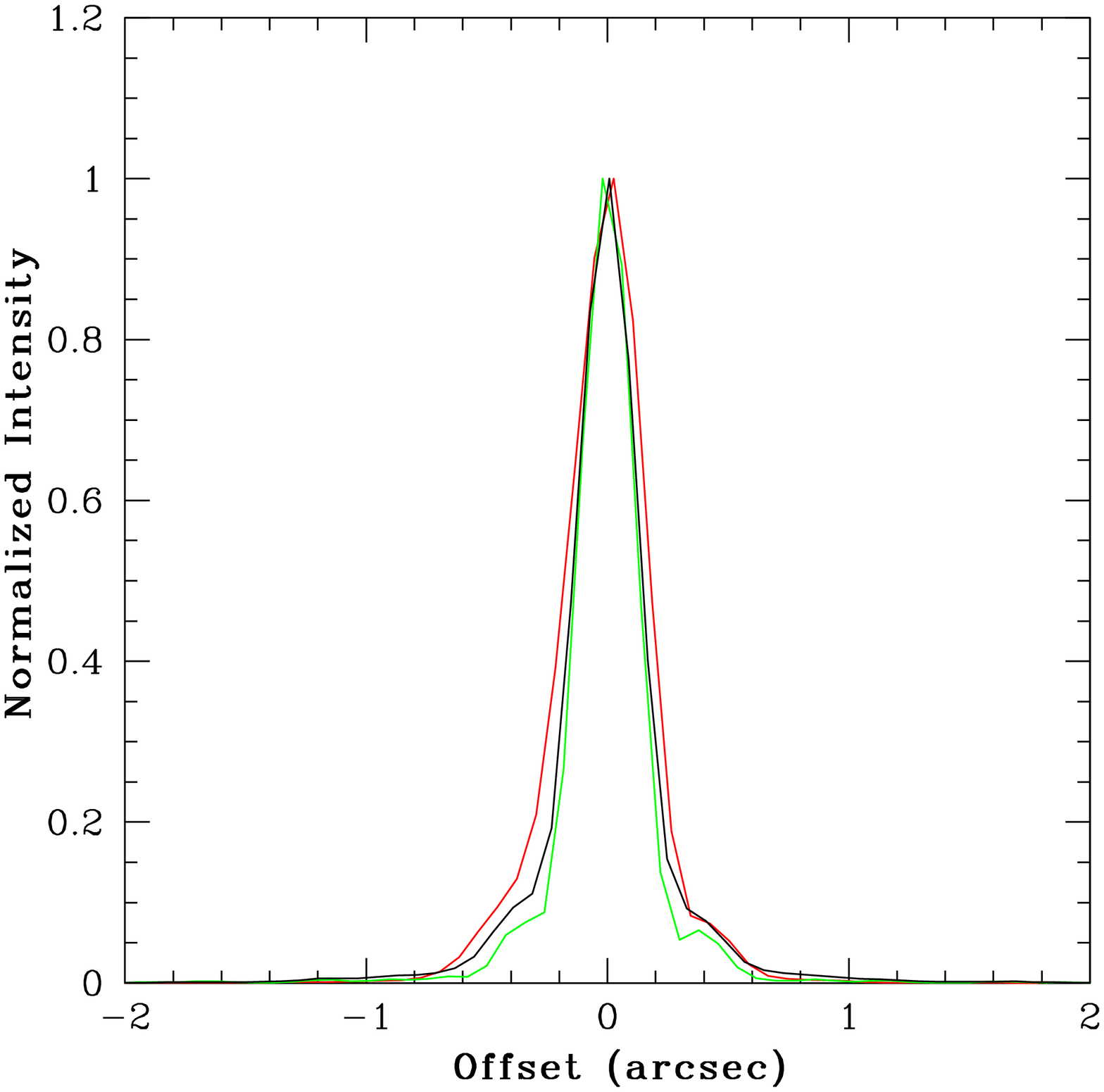} \\
\epsscale{1}
\plottwo{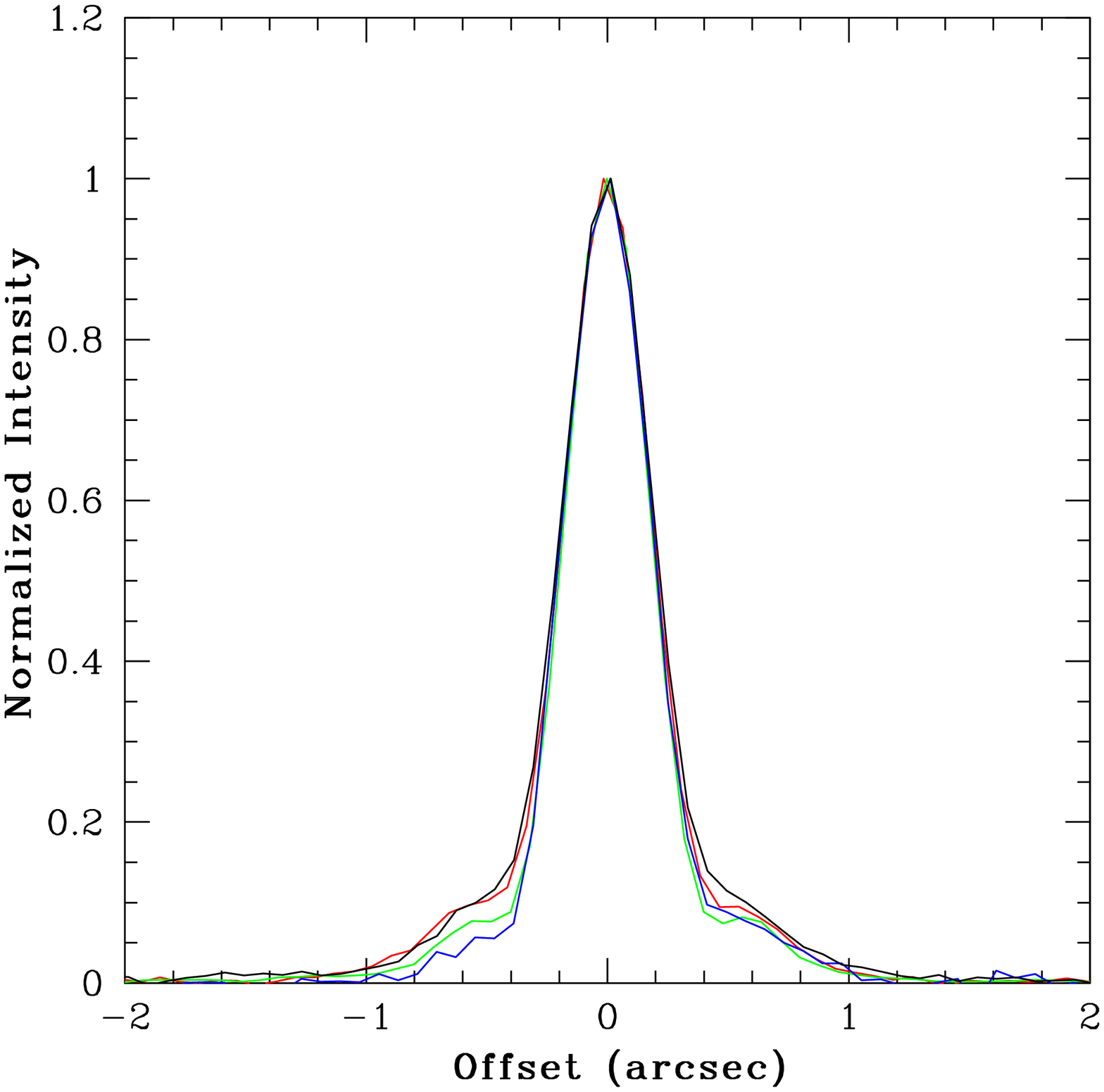}{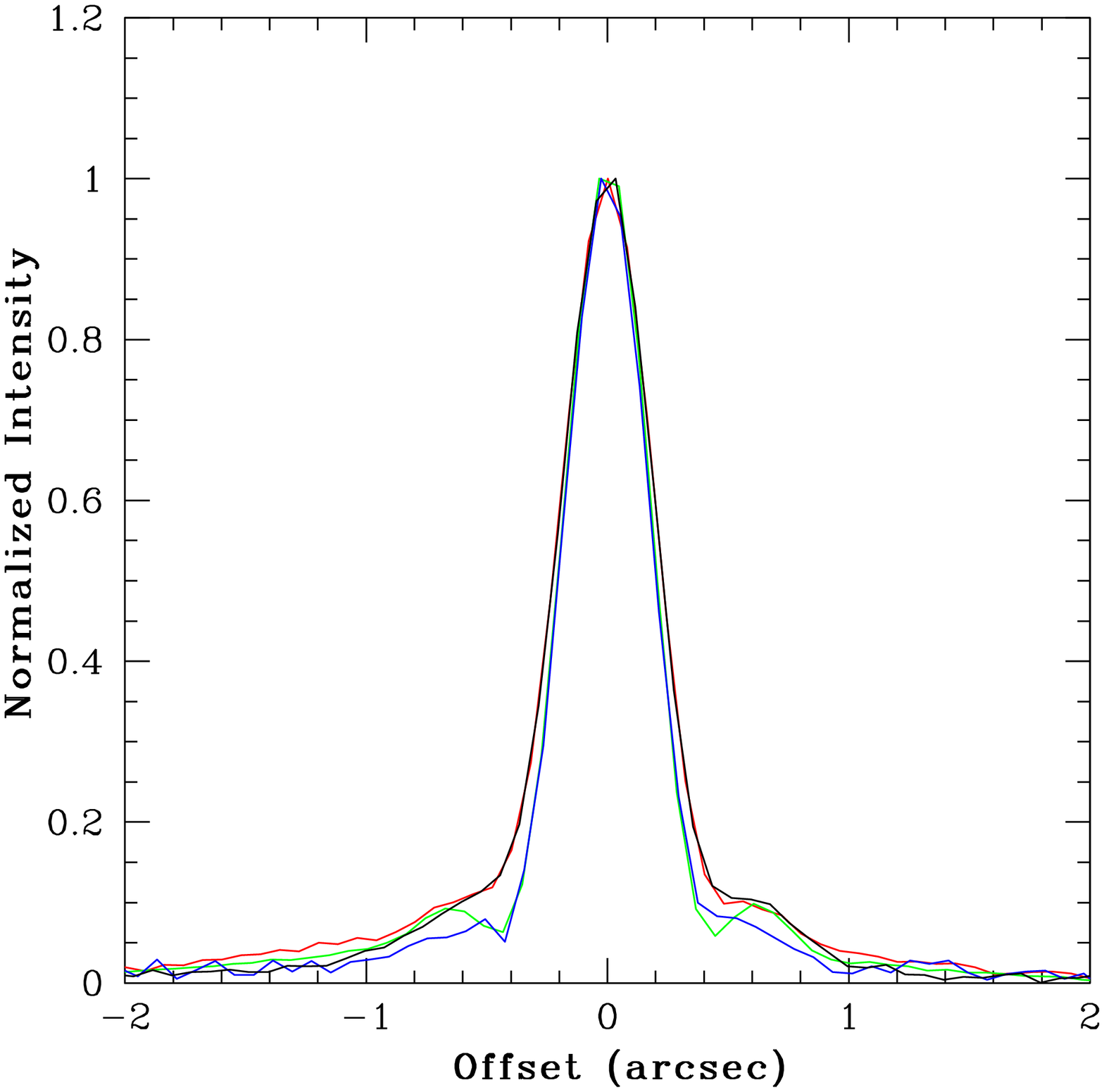}
\caption{(a) Line cuts through AB Aur (black) and our 11.7 $\mu$m standard 
stars $\gamma$ Dra (red) and $\alpha$ Cet (green). (b) Line cuts through 
AB Aur and our 18.7 $\mu$m standard stars $\gamma$ Dra (red), $\alpha$ Cet 
(green), and MWC 480 (blue) made perpendicular to the direction of the 
asymmetric scattered light. (c) Line cuts through AB Aur and our 18.7 $\mu$m 
standard stars $\gamma$ Dra (red), $\alpha$ Cet (green), and MWC 480 (blue) 
made parallel to the direction of the asymmetric scattered light.}
\end{figure}

\begin{figure}[ht]
\figurenum{2}
\epsscale{1.0}
\plotone{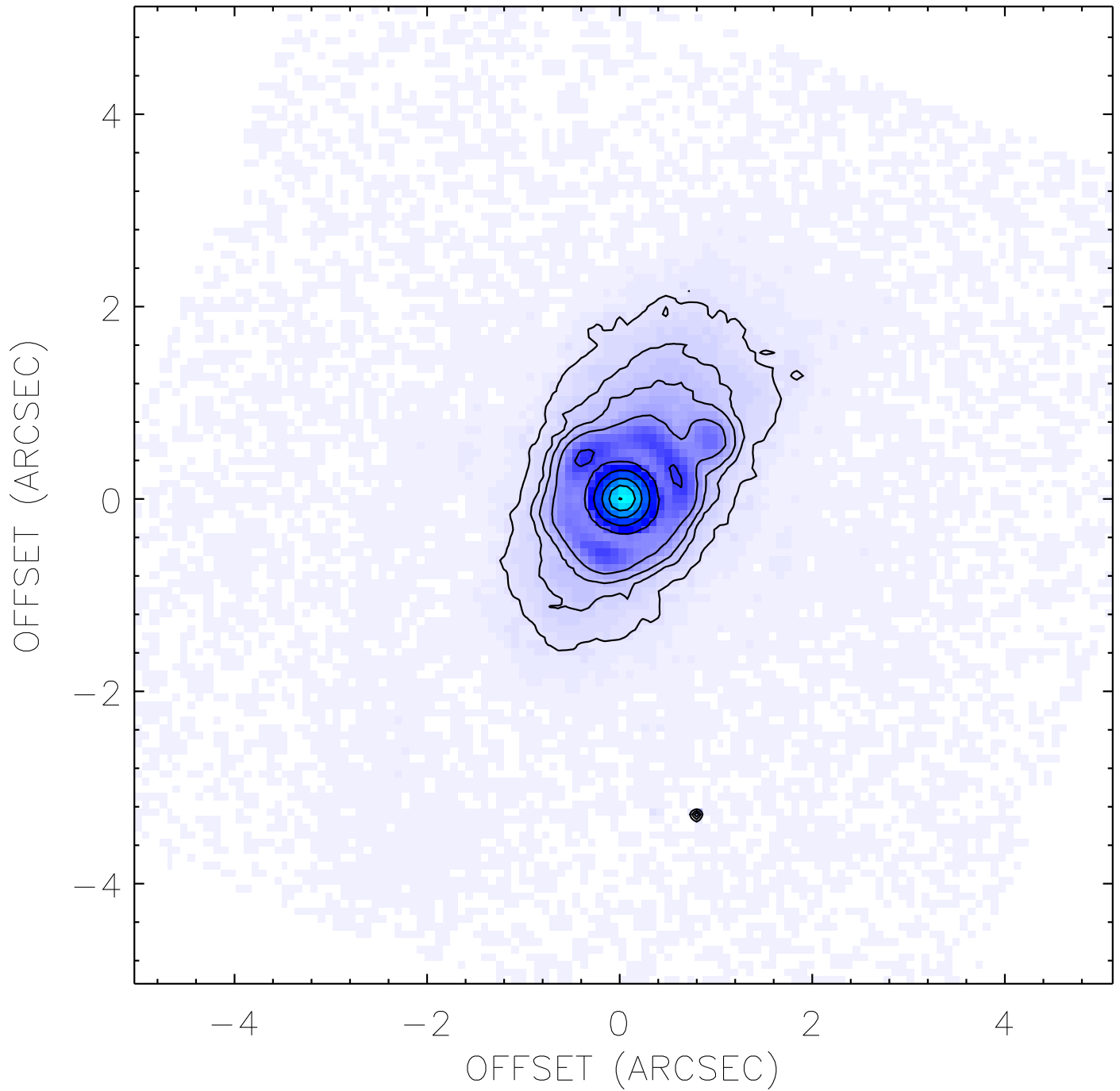}
\caption{The 18.7 $\mu$m image of $\alpha$ Cet, our standard star, rotated so 
that the ghost appears in the same position as in the image of AB Aur. The 
contours show 1\%, 2\%, 3\%, 5\%, 10\%, 25\%, 50\%, 75\%, and 99\% of the 
maximum surface brightness.}
\end{figure}

\begin{figure}[ht]
\figurenum{3}
\epsscale{1.0}
\plotone{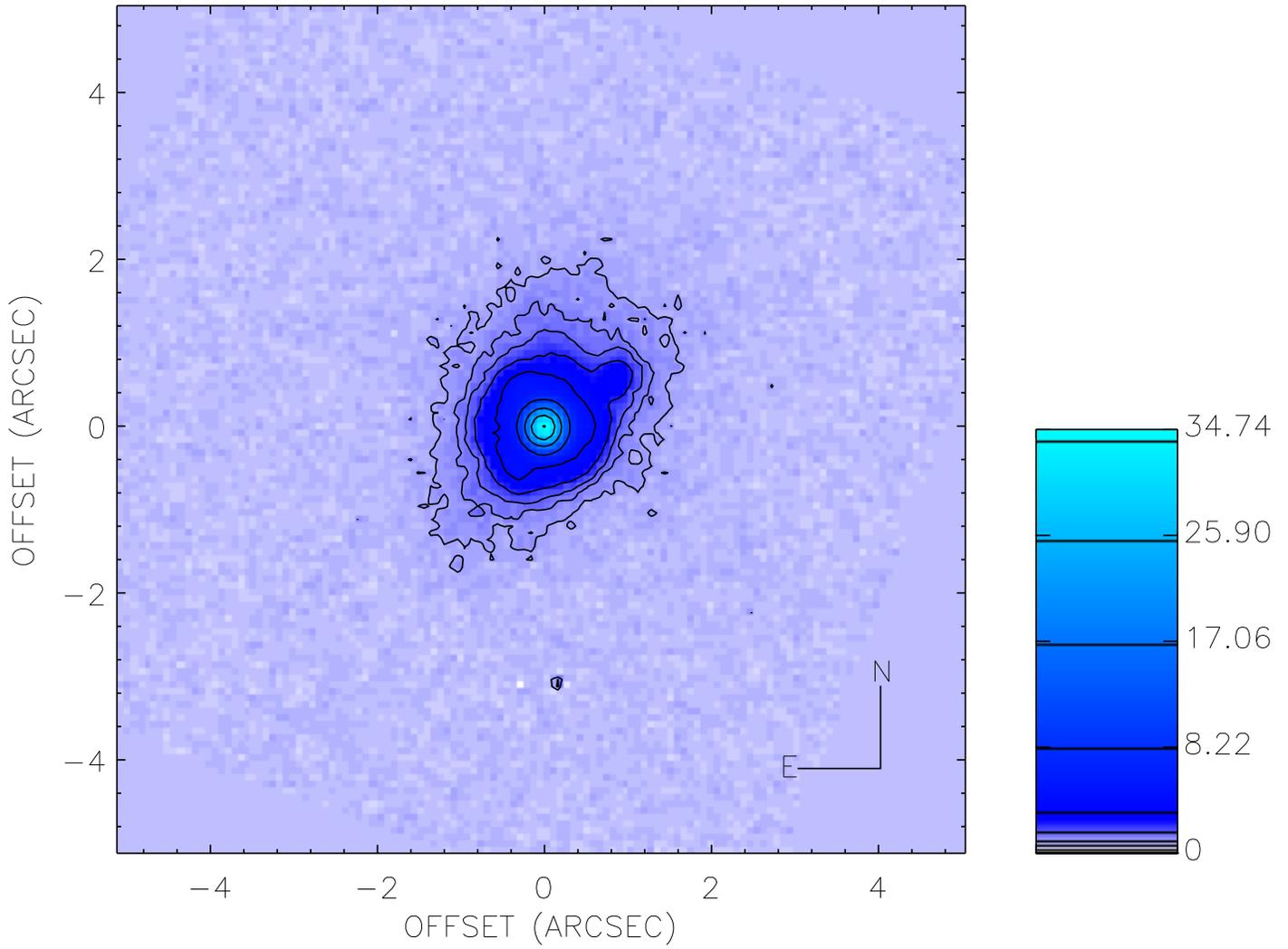}
\caption{The 18.7 $\mu$m image of AB Aur. North is up and East is to the left. 
The color bar shows the intensity key (Jy arcsec$^{-2}$). The contours show 
1\%, 2\%, 3\%, 5\%, 10\%, 25\%, 50\%, 75\%, and 99\% of the maximum surface 
brightness.}
\end{figure}

\begin{figure}[ht]
\figurenum{4}
\plottwo{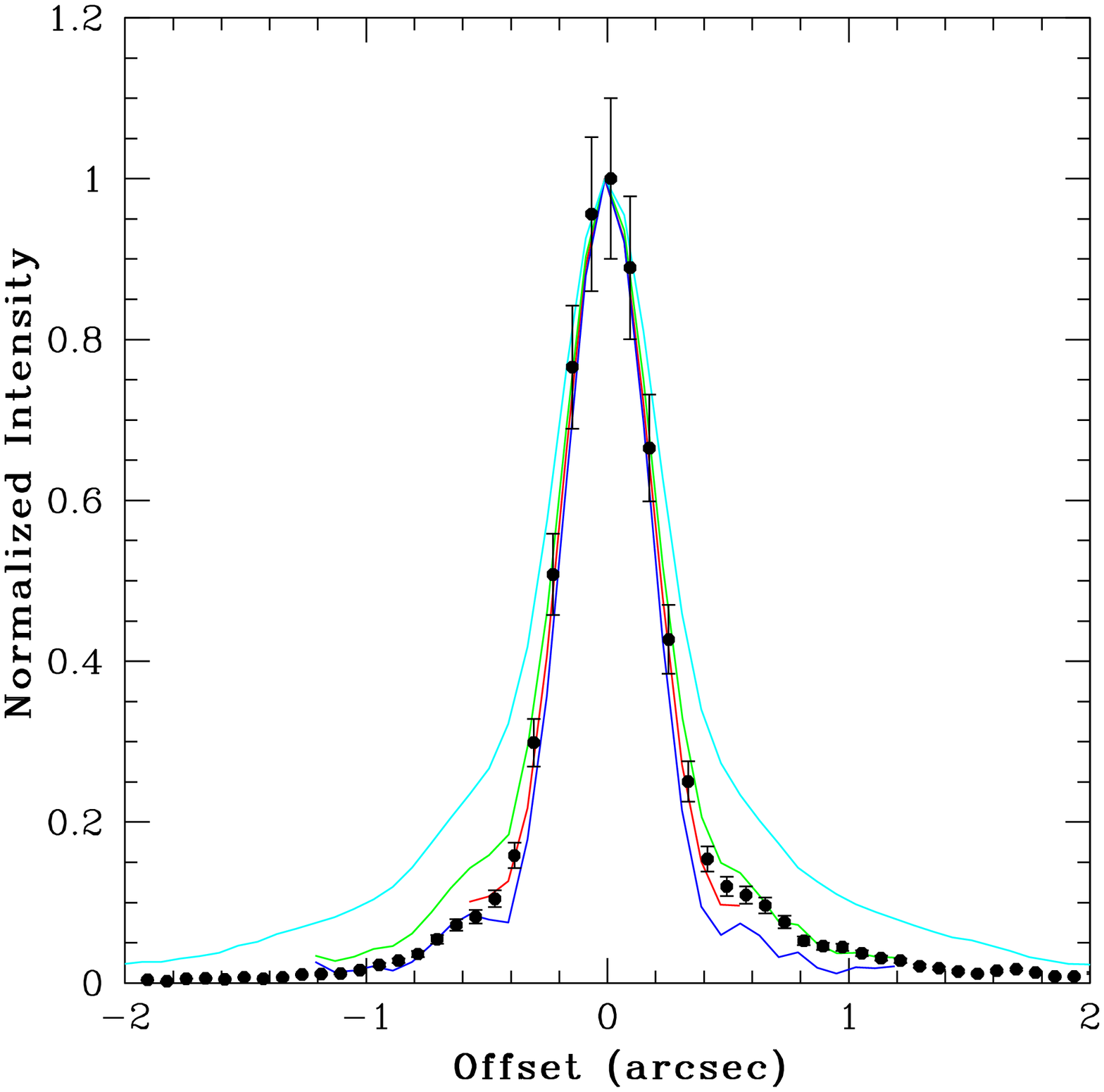}{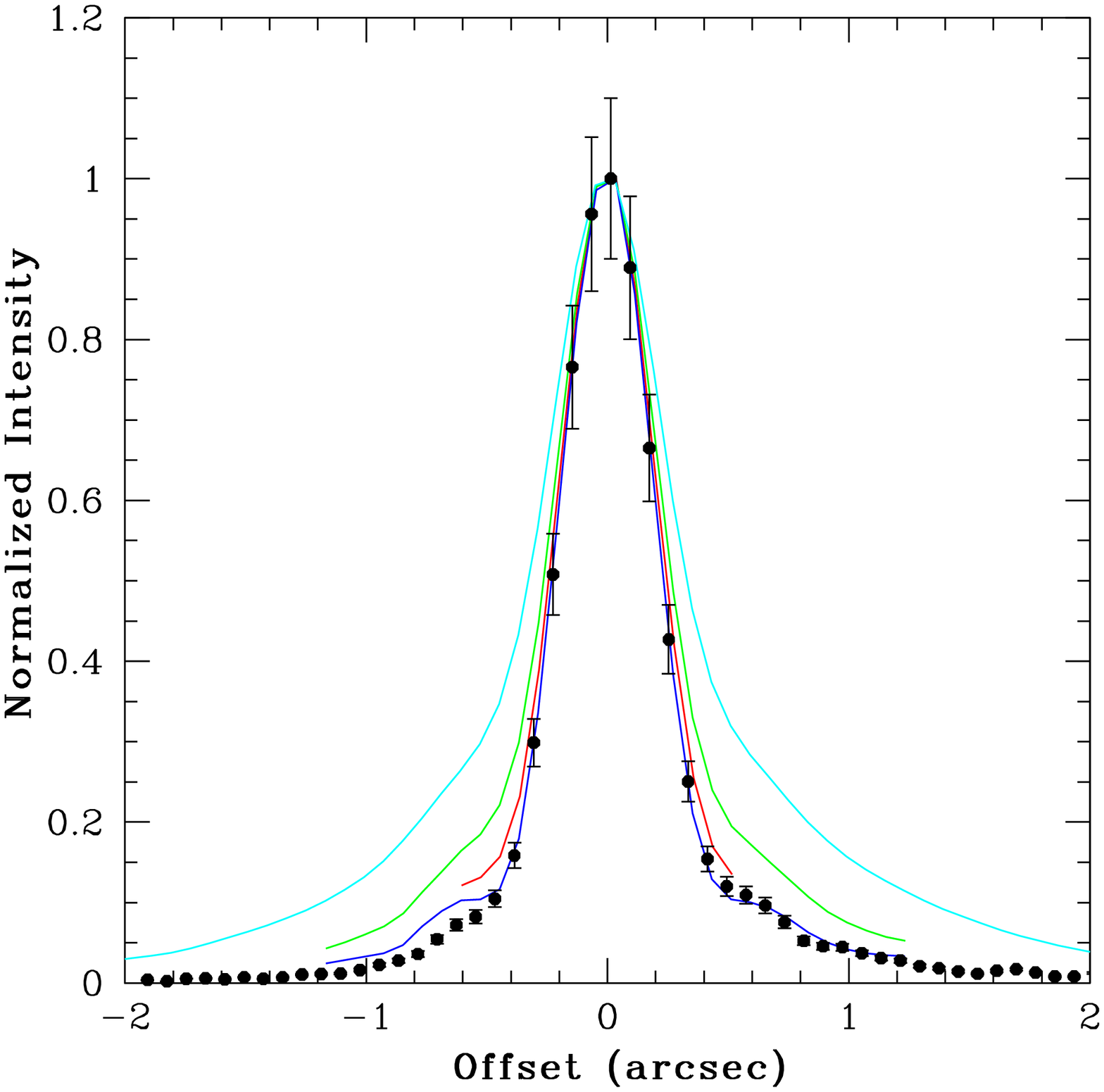}
\caption{(a) Line cuts North-South through AB Aur at 18.7 $\mu$m and each of 
the models convolved with MWC 480, the PSF observed closed in time to AB Aur.
(b) Line cuts North-South through AB Aur at 18.7 $\mu$m and each of the models
convolved with $\gamma$ Dra, the standard star observed with the worst seeing.
The blue, red, green, and cyan lines (from bottom to top )show the D'Alessio
(2002), Vinkovic et al. (2002), Bouwman et al. (2000), \& Dullemond et al. 
(2001) models respectively.}
\end{figure}

\end{document}